\documentclass[journal,11pt,draftcls,onecolumn]{IEEEtran}
\usepackage{array}
\usepackage{mdwmath}
\usepackage{mdwtab}
\usepackage{fixltx2e}
\usepackage{url}
\usepackage{color,soul}
\usepackage{xcolor}
\usepackage[]{graphicx}
\usepackage{epsfig}
\usepackage{bbm}
\usepackage{enumerate}
\usepackage{amsfonts}
\usepackage{amsmath}
\usepackage{amssymb}
\usepackage{cite}
\usepackage{algorithm}
\usepackage{algpseudocode}
\usepackage{eqparbox}
\usepackage{comment}
\usepackage{float}
\usepackage{makecell}
\newcolumntype{x}[1]{>{\centering\arraybackslash}p{#1}}
\usepackage{tikz}
\usepackage{mathrsfs}
\usepackage{array}
\usepackage{mdwmath}
\usepackage{mdwtab}
\usepackage{eqparbox}
\usepackage{float}
\usepackage{sidecap}
\usepackage{enumitem} 

\usepackage{blindtext}

\usepackage{subfiles} 

\providecommand{\norm}[1]{\lVert#1\rVert}

\DeclareMathOperator*{\argmin}{argmin}

\usepackage{lipsum}

\newcommand\blfootnote[1]{%
  \begingroup
  \renewcommand\thefootnote{}\footnote{#1}%
  \addtocounter{footnote}{-1}%
  \endgroup
}

\newif\ifarxiv
\arxivtrue

\begin{document}
\title{Algorithm-driven Advances for Scientific CT Instruments: From Model-based to Deep Learning-based Approaches}
\author{S.~V.~Venkatakrishnan{$^{\star}$}, K.~Aditya~Mohan{$^{\circ}$}, Amir~Koushyar~Ziabari{$^{\star}$}, and Charles A. Bouman{$^{+}$} \\
 \small{  $^{\star}$ Multimodal Sensor Analytics Group, Oak Ridge National Laboratory, Oak Ridge, TN 37831, USA.}\\
 \small{ $^{\circ}$ Computational Engineering Division, Lawrence Livermore National Laboratory, Livermore, CA 94551, USA.} \\
  \small{  $^{+}$ School of Electrical and Computer Engineering, Purdue University, West Lafayette, IN 47907, USA.} \\
   \small{ $^{\star}$\{venkatakrisv, ziabarik\}@ornl.gov, $^{\circ}$mohan3@llnl.gov, $^{+}$bouman@purdue.edu}
}
\maketitle
 \blfootnote{\tiny This manuscript has been authored by UT-Battelle, LLC, under Contract No. DE-AC05-00OR22725 with the U.S. Department of Energy. 
  The United States Government and the publisher, by accepting the article for publication,acknowledges  that  the  United  States  Government  retains  a  non-exclusive,paid-up,  irrevocable,  world-wide  license  to  publish  or  reproduce  the  published form of this manuscript,  or allow others to do so,  for United States Government purposes.   
  DOE will provide public access to these results of federally sponsored research in accordance with the DOE Public Access Plan(http://energy.gov/downloads/doe-public-access-plan).
  This work was performed under the auspices of the U.S. Department of Energy by Lawrence Livermore National Laboratory under Contract DE-AC52-07NA27344.
  This work was partially supported by Oak Ridge National Lab via the Artificial Intelligence Initiative. 
  S.V. Venkatakrishnan was partially supported by the U.S. DOE Office of Basic Energy Science.  
  A.K. Ziabari was supported by the U.S. DOE, Office of Energy Efficiency and Renewable Energy, Advanced Manufacturing Office, under contract DE-AC05-00OR22725 with UT-Battelle, LLC. This work was partially supported by NSF grant number CCF-1763896.
  }
\section{Introduction \label{sec:intro}}
\begin{figure}
\begin{center}
\includegraphics[scale=0.50,trim=0cm 8cm 0cm 0cm,clip]{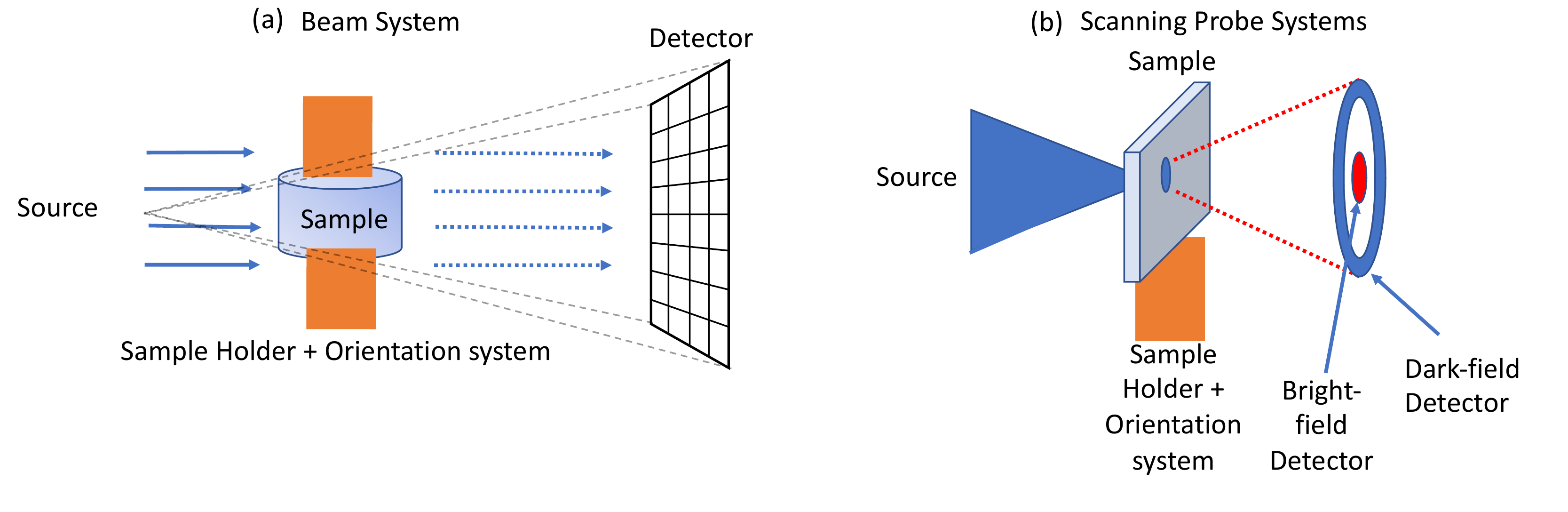} 
\end{center}
\vspace{-0.3in}
\caption{\label{fig:diff_inst} A schematic of common acquisition set-ups used in scientific CT systems - 
(a) is used in parallel-beam and cone-beam X-ray/neutron CT systems, (b) is used in  electron/X-ray microscope-based scanning probe systems systems. 
In each case a source is used to illuminate the sample of interest and a detector system (area detector, point detector, or annular detector) captures the result of this interaction. 
The sample, which is mounted on a holder, is re-oriented in order to make a collection of measurements. 
The archetypal acquisition geometry for SCT instruments is to rotate the sample about a single axis perpendicular to the direction of the incident source and make a collection of measurements followed by reconstruction using analytic algorithms. 
}
\end{figure}
Multi-scale 3D characterization is widely used by materials scientists to further their understanding of the relationships between microscopic structure and macroscopic function. 
Scientific computed tomography (CT) instruments are one of the most popular choices for 3D non-destructive characterization of materials at length scales ranging from the angstrom-scale to the micron-scale. 
These instruments typically have a source of radiation (electrons, X-rays, neutrons\ifarxiv
\cite{saghi2012electron,stock2008recent,ercius2015electron,singh20193d}
\fi) that interacts with the sample to be studied and a detector assembly to capture the result of this interaction (see Fig.~\ref{fig:diff_inst}). 
A collection of such high-resolution measurements are made by re-orienting the sample which is mounted on a specially designed stage/holder after which  \textit{reconstruction algorithms} are used to produce the final 3D volume of interest. 
The specific choice of which instrument to use depends on the desired resolution and properties of the materials being imaged. 
The end goal of scientific CT scans include determining the morphology, chemical composition or dynamic behavior of materials when subjected to external stimuli.
In summary, scientific CT instruments are powerful tools that enable 3D characterization across multiple length scales and play a critical role in furthering our understanding of the structure-function relationships of different materials. 

\section{Challenges in Scientific CT \label{sec:chal}}

\begin{figure}
\begin{center}
\includegraphics[scale=0.50,trim=0cm 0cm 0cm 0cm,clip]{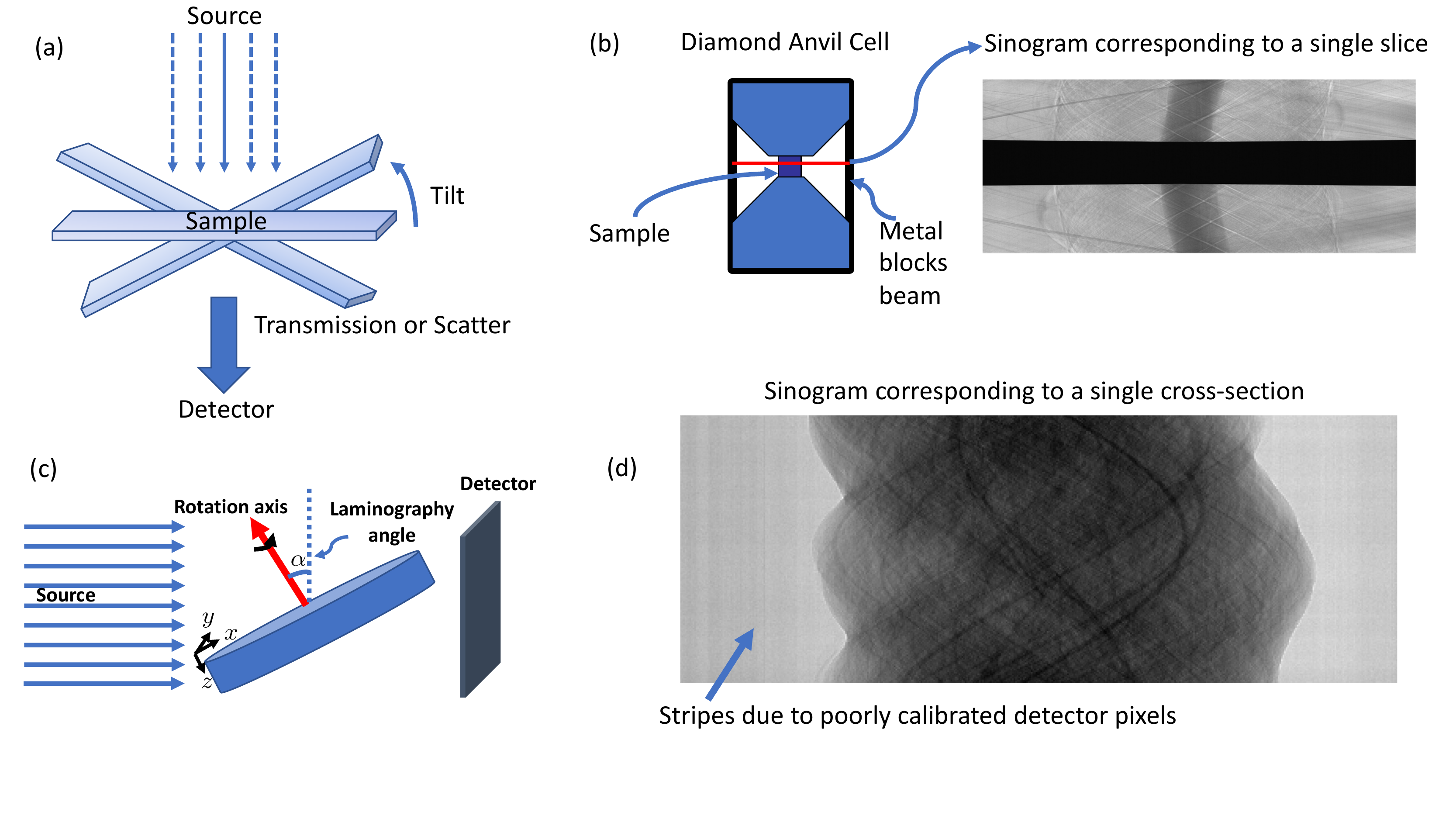} 
\end{center}
\vspace{-0.5in}
\caption{\label{fig:artifacts}  Illustration of some challenges in SCT. 
Due to mechanical limitations of the sample holder (a) and (b), the shape of the sample being imaged (c) and poorly calibrated detectors (d), it can be challenging to obtain accurate 3D reconstructions from  the resulting limited view, sparse, low-SNR data.
Note that a \textit{sinogram} refers to a particular organization of the measured data from a CT scan where the vertical axis corresponds to the orientation and the horizontal axis corresponds to a single row of the detector. 
}
\end{figure}
The archetypal form of CT involves illuminating a sample with a beam, measuring a projection image corresponding to the transmitted or scattered signal and collecting a set of measurements by rotating the sample (or the source-detector system) about a single axis in the $0^{\circ}-180^{\circ}$ or $0^{\circ}-360^{\circ}$ range followed by a reconstruction routine that inherently assumes a linear relationship between the measured signal (or some pre-processed version) and the quantity to be reconstructed. 
While these type of systems are common in medical X-ray CT, there are several aspects that make the scientific computed tomography (SCT) problem different and challenging.
These challenges can be grouped into a few broad categories: \\ 
    \textbf{Limited-angle measurements}:
    In applications such as electron tomography (one of the most popular methods for angstrom-scale and nano-scale 3D imaging), the mechanical limitations of the sample holder along with the unique shape of the samples  may only allow for acquiring data in a limited angular range  (+/- $60^{\circ}$) \cite{ercius2015electron} as illustrated in Fig.~\ref{fig:artifacts} (a).
    Limited-angle data sets can also occur in other SCT modalities when the sample holders are designed for specialized tasks. 
    An example of this type of holder is the diamond anvil cell \cite{venkatakrishnan2016making} (used for studying the properties of materials under extremely high pressure) which strongly attenuates the incident beam in certain orientations (see Fig.~\ref{fig:artifacts} (b)). 
    In summary, driven by the flexibility required to engineer sample holders in order to perform novel  experiments, \textit{limited-angle} data-sets can occur in important SCT applications making it challenging to obtain high-quality reconstructions. \\
    \textbf{Unconventional measurement geometries}: The acquisition geometry used in SCT instruments can be different compared to the conventional single-axis CT setup. 
    The need for new geometries is driven by the requirement to measure samples with unique shapes which may not yield sufficient signal-to-noise ratio (SNR) data in certain orientations. 
    For example, a technique called laminography has been developed to measure samples that are lamellar  such as integrated circuit boards.  
    In laminography 
    \ifarxiv
    \cite{salvemini2015neutron,VenkatLam17}, 
    \else
    \cite{VenkatLam17}
    \fi
    the sample is tilted and rotated about this new tilt-axis in order to  measure a signal of sufficient strength on the detector (Fig.~\ref{fig:artifacts} (c)). 
    Unlike in the case of conventional single-axis CT, analytic reconstruction algorithms for novel geometries are not readily available thereby impeding the use of novel acquisition schemes. 
    In summary, the unique shapes of various samples to be scanned dictate a greater degree of flexibility in the measurement geometries for SCT instruments and require novel reconstruction algorithms.  \\
    \textbf{Sparse, low SNR, and poorly calibrated data}:
    SCT instruments are typically purchased from commercial vendors or are built at scientific user facilities (SUF) where a source of radiation/particles (a high-flux of neutrons from a nuclear reactor, monochromatic X-rays from a synchrotron etc.) forms the basis for a unique imaging capability. 
    In both situations, SCT instruments are often very expensive and are treated as a shared resource; leading to a need for making the fewest possible measurements in order to extract the relevant scientific information from the study. 
    As a result, reducing the number of measurements (sparse view data) and the duration of each measurement (leading to low SNR) can be critical to maximize the throughput in order to make SCT instruments available to a large number of users. 
    Sparse-view and low-SNR data can also occur in the SCT experiments where the sample can suffer radiation damage as with the case of bio-materials.
    This type of data is also common in high-speed time-resolved 4D-CT experiments \cite{AdityaTIMBIR}, where the goal is to image how a sample is changing at the microscopic scales when subjected to external stimuli. 
    
    SCT measurements can also be corrupted by different signals that are \text{independent} of the sample. 
    For example, it is common in conventional X-ray micro-CT \cite{AdityaICASSP14} and neutron CT \cite{VenkatLam17} 
    systems
    to have spurious radiation strike the detector leading to a high-amplitude signal in a few measurements. 
    Furthermore, the detectors used in SCT instruments may not be perfectly calibrated.
    One common example of this phenomenon is the observation of correlated ``streaks'' in the measured sinograms (a way of organizing the CT data so that the data corresponding to all orientations for a single slice can be easily visualized) because the gain associated with each detector pixel is different (see Fig.~\ref{fig:artifacts} (d)). 
    These imperfections in the data due to outliers and poorly calibrated detectors results in reconstructions with streak and ring artifacts when a conventional reconstruction algorithm is directly applied to the data. 
    In summary, it is challenging to achieve higher throughput, reduce damage to samples by lowering their exposure to source-radiation, and improve the spatio-temporal resolution of 4D-CT while preserving image fidelity because of the sparse, low-SNR and poorly calibrated measurements. \\
    \textbf{Large data-sets}:
    SCT scans are usually conducted in order to obtain 3D information at high resolutions. 
    With the advent of faster, higher pixel resolution detectors and the 
    need to measure
    larger samples, there has been an explosion in the size of SCT data-sets.
    For example, it is common across SCT applications to use detectors which are approximately $2000 \times 2000$ pixels, and corresponding CT reconstructions to have sizes of the order of $2000 \times 2000 \times 2000$ voxels.
    In the case of hyper-spectral SCT instruments \cite{gursoy2015hyperspectral,venkat2021hsct}, the size of the data is even larger depending on the number of hyper-spectral channels. 
    For 4D-CT, this problem is compounded since  the number of measurements increase linearly with time.
    In summary, it can be challenging to obtain high-quality reconstructions in reasonable time-frames for SCT applications. 
    
\section{Conventional Approaches to SCT}
Despite of significant advances made in developing various hardware components of SCT instruments (source, lenses, sample holders, detectors, etc.), until recently there has been less focus on the development of reconstruction algorithms in order to deal with the various challenges encountered.  
A common practice has involved measuring a large amount of data corresponding to the Nyquist criterion \cite{KakSlaney} or the maximum number of measurements at reasonable SNR that can be made in an allocated amount of time in the case of a shared instrument at SUFs. 
Following the acquisition, the measurements are 
pre-processed (filters to suppress outliers, heuristic  correction of mis-calibrated data, normalization) and reconstructed using analytic algorithms  such as the filtered back projection (FBP) \cite{kak2002principles}, gridrec \cite{marone2012regridding} or Feldkamp-Davis-Kreiss (FDK) \cite{feldkamp1984practical} because of their widespread availability and low computational complexity. 
However, the performance of these algorithms can be poor when dealing with non-linearities in the measurement, the presence of high levels of noise, and the limited number of measurements - which are common in the context of scientific CT applications as discussed in Section~\ref{sec:chal}.  
The reliance on the use of analytic reconstruction techniques in turn limits the characterization capability of SCT instruments by resulting in significant artifacts from the sparse-view, limited, and low-SNR data-sets. 
Additionally, the reliance on analytic reconstruction algorithms has led to an inefficient usage of the instruments by requiring the collection of large amounts of data in order to ensure the reconstructed images are of high-quality. 

In the rest of this article, we will present an overview of recent advances in 
non-linear
reconstruction algorithms that have enabled significant improvements in the performance of scientific CT instruments - enabling faster, more accurate and novel imaging capabilities. 
We emphasize that while this article focuses on scientific CT applications where a linear forward model accurately describes the physics of image formation (up to to point wise normalization), there are important CT applications such a phase-contrast imaging
\ifarxiv
\cite{Burvall2011PRSurvey} 
\fi
and ptychography 
\ifarxiv
\cite{pfeiffer2018x} 
\fi
where the underlying physics-based model is significantly more complicated, but for which the ideas presented here are equally relevant.
In the first part, we will focus on model-based image reconstruction (MBIR) algorithms \ifarxiv
\cite{Bo641Text,fessler2010model} 
\else
\cite{Bo641Text} 
\fi
that formulate the inversion as solving a high-dimensional optimization problem involving a data-fidelity term (which includes a physics-based forward model) and a regularization term (based on a model for the sample to be imaged). 
By accurately modeling the physics and noise statistics of the measurement and combining it with state-of-the art regularizers, we will highlight how dramatic improvements are being made in the performance of several types of scientific CT instruments.
While the development of MBIR methods have demonstrated that it is possible to dramatically improve the performance of CT instruments, these methods are computationally expensive for the high-resolution scans encountered in scientific CT applications.
This bottleneck had led researchers to adapt and develop non-iterative deep-learning (DL) approaches based on convolutional neural networks \cite{mccann2017convolutional} to attain similar improvements as the MBIR methods in certain scenarios.
In the last part of the article, we will present an overview of recent approaches using DL based algorithms for improving scientific CT instruments.
We will summarize different approaches developed in order to address the tomographic inversion - including  data-domain learning, and image-domain learning. 
The recent advances have shown that DL-based methods are a promising tool to complement MBIR methods because of their rapid inference time on large high resolution scientific CT data sets while enabling similar improvements in image quality, and reduction of the scan time.    
The rest of this article is organized as follows. 
In section \ref{sec:mbir}, we present the MBIR framework and discuss how it has been adapted to address challenges in 3D-CT (section \ref{sec:mbir_3d}), and 4D-CT (section \ref{sec:mbir_4d}) including the development of computational techniques to handle large data sets (section \ref{sec:mbir_acc}). 
In section \ref{sec:dl}, we will present a survey of different DL techniques that have been developed for scientific CT systems and in section \ref{sec:concl} we present concluding statements.

\section{Model-based Image Reconstruction for Scientific CT Instruments \label{sec:mbir}}
Model-based image reconstruction (MBIR) 
\ifarxiv
\cite{Bo641Text,fessler2010model} 
\else
\cite{Bo641Text} 
\fi
refers to an umbrella term for joint maximum \textit{a posteriori} (MAP) estimation \cite{DjafariJointMAP}  or a regularized inversion approach to solving image reconstruction problems. 
In the MBIR framework, the reconstruction task is formulated as 
\begin{eqnarray}
\left(\hat{x},\hat{\psi}\right) \leftarrow \argmin_{x \in \Omega, \psi \in \Psi} \left\{ l(y;x,\psi) + r(x;\beta)\right\}
\label{eq:mbir_master}
\end{eqnarray}
where $y$ is a vector containing the measurements, $x$ is a vector corresponding to the object to be reconstructed, $l$ is a data-fidelity function that enforces consistency of the reconstruction with the measured data based on a physics-based forward model, $\psi$ is a vector of calibration parameters associated with the measurement, $\Omega$,  $\Psi$ are constraint sets, and  $r$ is a regularization term with parameters $\beta$.
In the context of MAP estimation \cite{Bo641Text}, the $l$ corresponds to the negative log-likelihood function and $r$ corresponds to the negative log-prior function. 
MBIR approaches have been used for several imaging problems and have enabled significant dose-reduction in medical X-ray CT 
\ifarxiv
\cite{YamadaLowDose12} 
\fi
and accelerations of MRI scans 
\ifarxiv
\cite{doneva2020mathematical} 
\fi
while preserving image quality compared to conventional approaches in the respective fields. 
The main challenges in the design of MBIR methods is the formulation of the cost-function of the type in equation~\eqref{eq:mbir_master} by an appropriate choice of the physics-based forward model, noise dependent data-fidelity loss, $l$, application dependent regularizer, $r$, and the design of fast optimization algorithms to obtain a minimum of the cost function. 
In the next three sections, we will present how different MBIR algorithms have been developed for 3D and 4D SCT.

\subsection{Volumetric CT \label{sec:mbir_3d}}
The goal of volumetric CT is to reconstruct some property of a sample such as linear attenuation coefficient, scatter coefficient, or complex valued index of refraction in 3D. 
The most straight forward adoption of MBIR for SCT has been for conventional transmission or scatter type CT, using a data-fidelity term of the form,
\begin{eqnarray}
l(y;x) = \frac{1}{2}\norm{y-Ax}_{W}^{2}
\label{eq:3dforward}
\end{eqnarray}
where $W$ is a diagonal matrix containing the inverse noise variance in the measurements, $A$ is the tomographic projection operator and $y$ either contains the log-normalized transmission measurements \cite{VenkatBF15} or the measured signal itself \cite{VenkatHAADF13} from each orientation. 
This model can be derived by assuming that the measurements are corrupted by additive white Gaussian noise or by using a quadratic approximation to the log-likelihood function based on Poisson statistics \cite{SaBo92}. 
A variety of regularizers have been combined with the model in \eqref{eq:3dforward}, but one popular class is the generalized Markov Random field-based (MRF) regularizer \cite{SaBo92} which includes the popular anisotropic total-variation and the q-generalized Gaussian MRF (qGGMRF) \cite{JBSaBoHsMultiSlice}. 
These regularizers are of the form,  
\begin{eqnarray}
r(x;\beta_s) = \beta_s \displaystyle \sum_{\{i,j\} \in \chi} w_{ij}\rho(x_{i} - x_{j})
\label{eq:mrf_prior}
\end{eqnarray}
where $\rho$ is a function that penalizes differences between neighboring voxels, $\beta_s$ is a parameter that adjusts the weight assigned to the  regularization terms, $\chi$ is a set containing all pairs of neighboring voxels in 3D and $w_{ij}$ are weights associated with each pair of voxels.    
MBIR algorithms based on combining models in \eqref{eq:3dforward} and \eqref{eq:mrf_prior} have been developed for parallel beam electron tomography
\ifarxiv
\cite{LevineBayesian,saghi2012electron,VandenBroek2012,GorisCompSenseET,MidgleyCompSenseET,VenkatHAADF13,VenkatBF15}, 
\else
\cite{LevineBayesian,saghi2012electron,VenkatHAADF13,VenkatBF15}
\fi
synchrotron based X-ray CT \cite{AdityaICASSP14,venkatakrishnan2016making}, and neutron tomography 
\ifarxiv
\cite{abir2016sparse,barnard2018total} 
\else
\cite{abir2016sparse}
\fi
enabling significantly higher quality reconstructions compared to the analytic reconstruction algorithms from sparse, limited-view and low-SNR data routinely encountered in these applications.
The development of MBIR methods has shown that it is possible to achieve similar image quality as the analytic reconstruction methods using about one-half or even one-fourth the typical number of measurements made \cite{venkatakrishnan2016making,abir2016sparse} at X-ray and neutron-CT instruments  - thereby potentially enabling two-four times more samples to be measured at these instruments than would have been possible when analytic reconstruction methods were used. 
Another advantage of using the model in \eqref{eq:3dforward}, is that the  $W$ matrix can be used as a simple means to adjust the relative weight of each measurement in order to reject sub-sets of measurements that are corrupted. 
Prior to the development of MBIR methods, the standard practice in several SCT applications has been to leave out entire projection data corresponding to a specific orientation due to the corruption of a sub-set of the data, resulting in an inefficient use of the measurements. 
For example in \cite{venkatakrishnan2016making}, the effective use of the weight matrix helped suppress streak artifacts due to beam-blocking caused by the strong attenuation from a diamond-anvil cell sample holder (see Fig.~\ref{fig:artifacts} (b)) in a CT study about the behavior of materials under extremely high-pressure. 
Finally, we note that while the quadratic data-fidelity term in
\eqref{eq:3dforward} has been demonstrated to be useful across several SCT applications, alternate models derived by assuming the measurements have a Poisson distribution that have the form 
\begin{eqnarray}
l(y;x)=\displaystyle \sum_{i=1}^{M} \{\left[Ax\right]_i - y_{i}\log(\left[Ax\right]_i) \}
\end{eqnarray}
where $M$ is the total number of measurements, have also been used along with regularizers of the form in \eqref{eq:mrf_prior} for low-dose SCT applications \cite{gursoy2015hyperspectral}. 

The MBIR approach using the models in~\eqref{eq:3dforward} and \eqref{eq:mrf_prior} has also been developed for novel computed tomography geometries.  
One such example is for \textit{laminography}, where the sample is rotated about a tilted axis instead of the conventional perpendicular axis (see Fig.~\ref{fig:artifacts} (c)) in order to image samples that might otherwise heavily attenuate the beam thereby requiring a very long scan time.
The authors in \cite{VenkatLam17} have developed an MBIR technique based on a new forward model term that incorporates the new acquisition geometry into the $A$ matrix in \eqref{eq:3dforward}. 
In addition to the benefits of enabling high-quality reconstructions from sparse-view and low-SNR data, MBIR approaches are useful for such novel geometries because analytic reconstruction algorithms may not be readily available or can result in significant artifacts in the reconstructed images. 
Similar MBIR approaches have also been recently developed for single particle cryo-electron microscopy 
\ifarxiv
\cite{donati2018fast,zehni2020joint,venkatcryo20}, 
\else
\cite{donati2018fast}, 
\fi
a widely used angstrom-scale 3D bio-imaging technique, in which the reconstruction involves inverting ultra low-dose data from parallel-beam projection images corresponding to \textit{arbitrary orientations} of the sample defined by a set of Euler angles.
This line of research has demonstrated that it is possible to obtain high-quality 3D reconstructions by using MBIR techniques despite the complicated geometry of acquisition associated with the SCT instrument, thereby allowing scientists to be able to have an additional control variable for their experimental acquisition.

Another powerful advantage of using MBIR methods for SCT is the ability to account for unknown calibration parameters associated with the measurement. 
In cases where the measurements are impacted by poorly calibrated detectors, conventional algorithms can produce reconstructions with significant artifacts. 
In order to address poorly calibrated data, the authors in \cite{VenkatHAADF13,AdityaICASSP14} modified the forward model in \eqref{eq:3dforward} to account for parameters such as detector gains and offsets. 
For example, in dark-field electron tomography \cite{VenkatHAADF13}, the gains and offsets of the detector are typically not measured. 
In order to address this challenge, a forward model of the form
\begin{eqnarray}
l(y;x,I,d)=\frac{1}{2}\norm{y-IAx-d}_{W}^{2}
\label{eq:haadf}
\end{eqnarray}
was proposed, where $I$ is a diagonal matrix containing the unknown gain associated with the detected signal at each projection orientation, and $d$ is a vector containing the unknown offsets. 
Using this model along with constraints on $I$ , resulted in an algorithm that significantly improved image quality compared to the traditional FBP method that was widely used in the field (see Fig.~\ref{fig:WBPvSIRTvBayes}). 
A similar approach \cite{AdityaICASSP14} was used to address the challenge of poorly calibrated detectors where each pixel in the detector has a different gain (as shown in Fig.~\ref{fig:artifacts} (d)), by modeling the unknown detector gains into the MBIR framework, leading to reconstructions that significantly suppress the ring artifacts that commonly result from such mis-calibrations (see Fig.~\ref{fig:MBIRvsGridrec}). 

\newcommand{\imagesize}{1.7in}
\begin{figure}[!htbp]
\begin{center}
\begin{tabular}{ccc}
MBIR Volume Rendering & FBP (analytic reconstruction) & MBIR
\tabularnewline
\mbox{\epsfig{figure=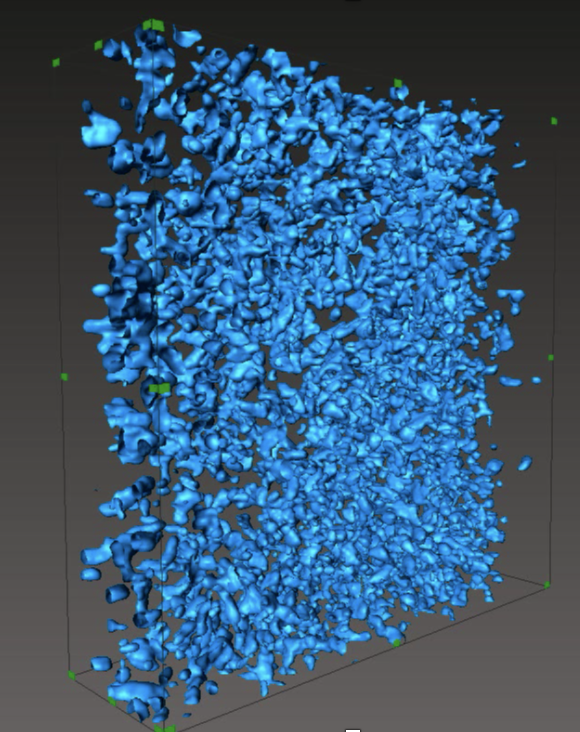,width=1.2in}} & 
\mbox{\epsfig{figure=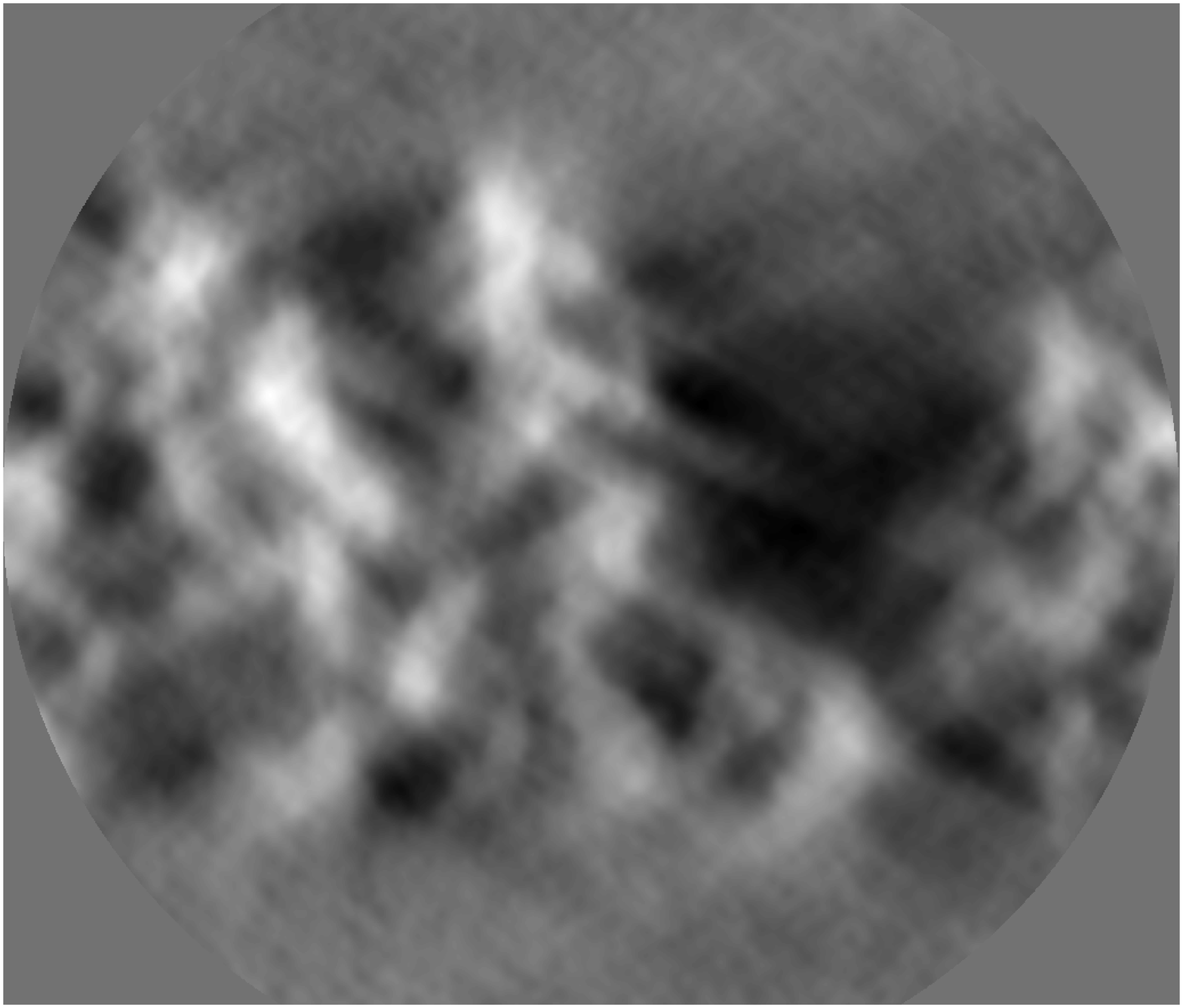,width=\imagesize}} & 
\mbox{\epsfig{figure=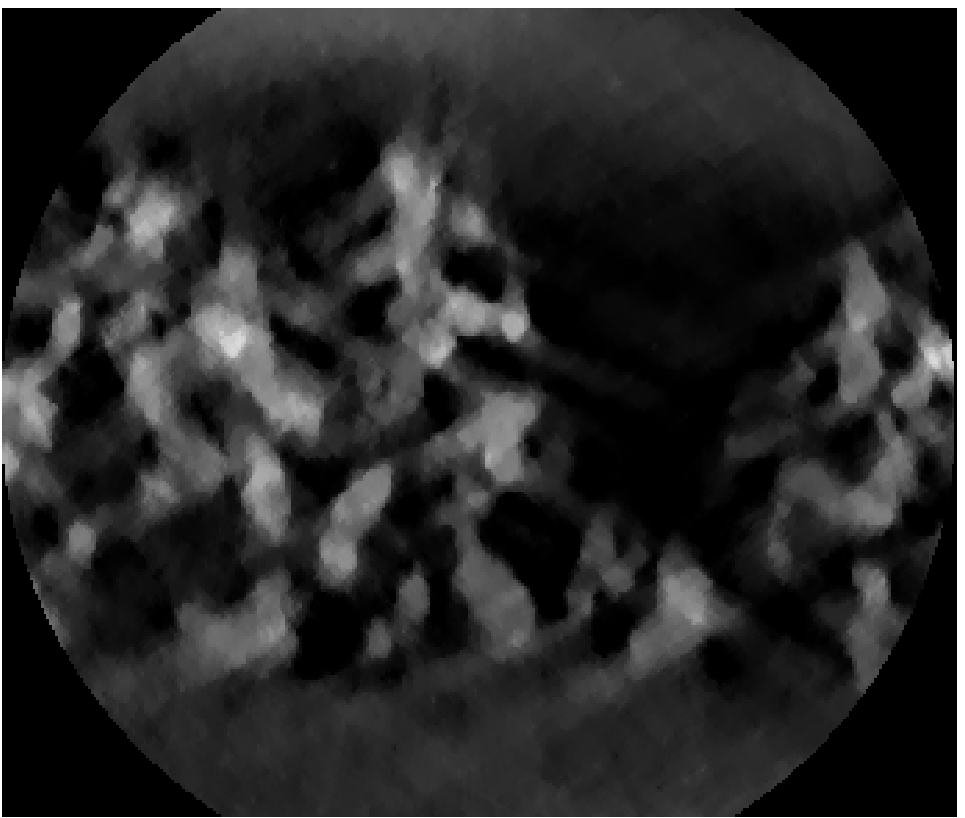,width=\imagesize}}
\end{tabular}
\end{center}
\caption{\label{fig:WBPvSIRTvBayes} Comparison of MBIR with FBP on an experimental high-angle annular dark field electron tomography data set of Titanium di-oxide nano-particles (adapted with permission from \cite{VenkatHAADF13}). 
The illustration includes a 3D rendering and a single cross-section from the 3D reconstruction obtained using FBP and the MBIR approach. 
The data-set contained $60$ projection images of size $1024 \times 1024$ pixels measured in an angular range of $+/- 60^{\circ}$.  
In spite of the low SNR, sparse, limited-view data with unknown calibration parameters, the MBIR method significantly suppresses artifacts compared to the FBP method. 
This highlights how the use of powerful reconstruction algorithms can improve the imaging capability of SCT instruments. 
}
\end{figure}
\newcommand{\imwidth}{2.5in}
\begin{figure}[!htbp]
\vspace{-0.025in}
\begin{center}
\begin{tabular}{cc} 
Gridrec (analytic reconstruction)  & MBIR  \\
\mbox{\epsfig{figure=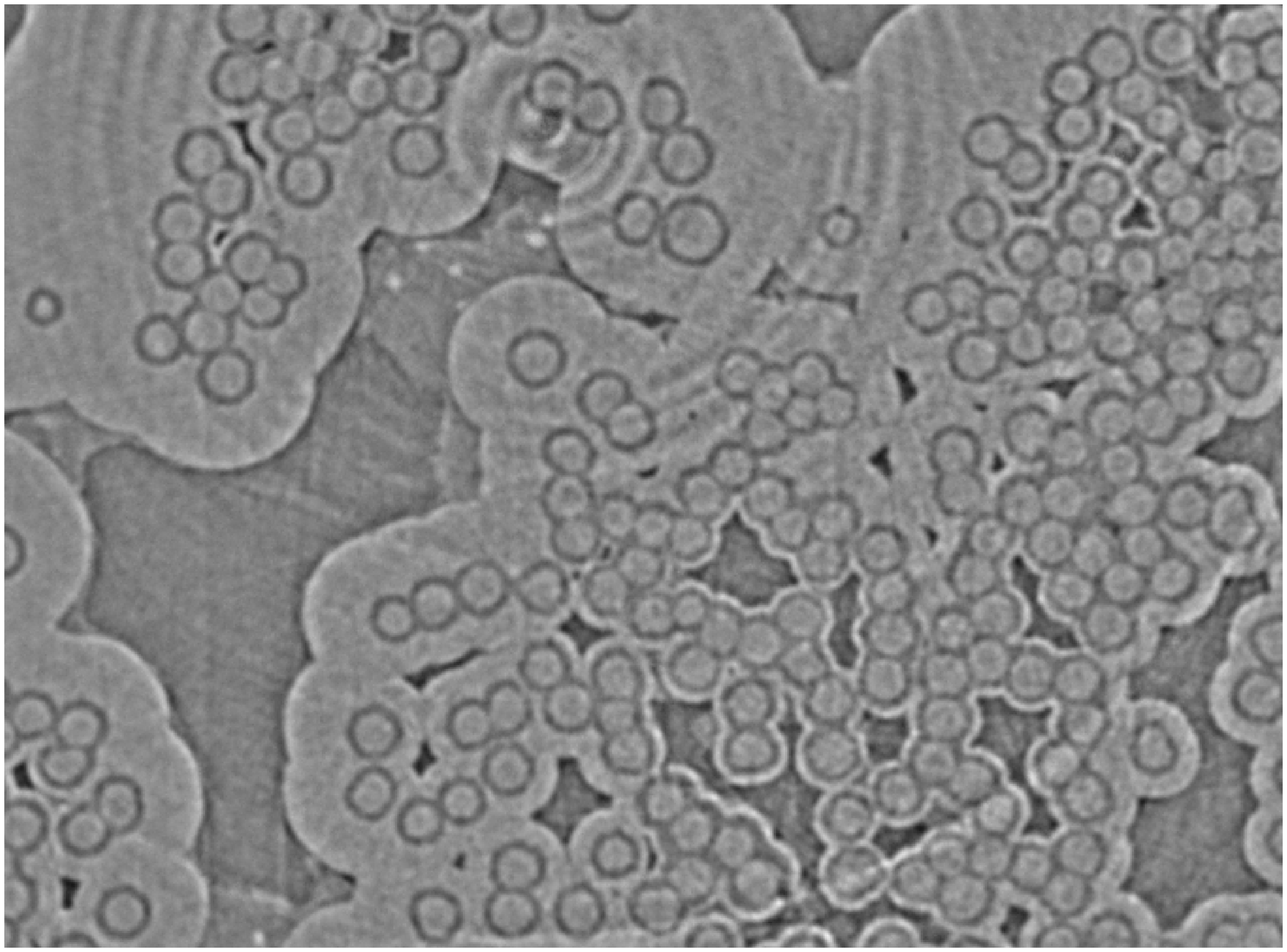, width=\imwidth}}  &
\mbox{\epsfig{figure=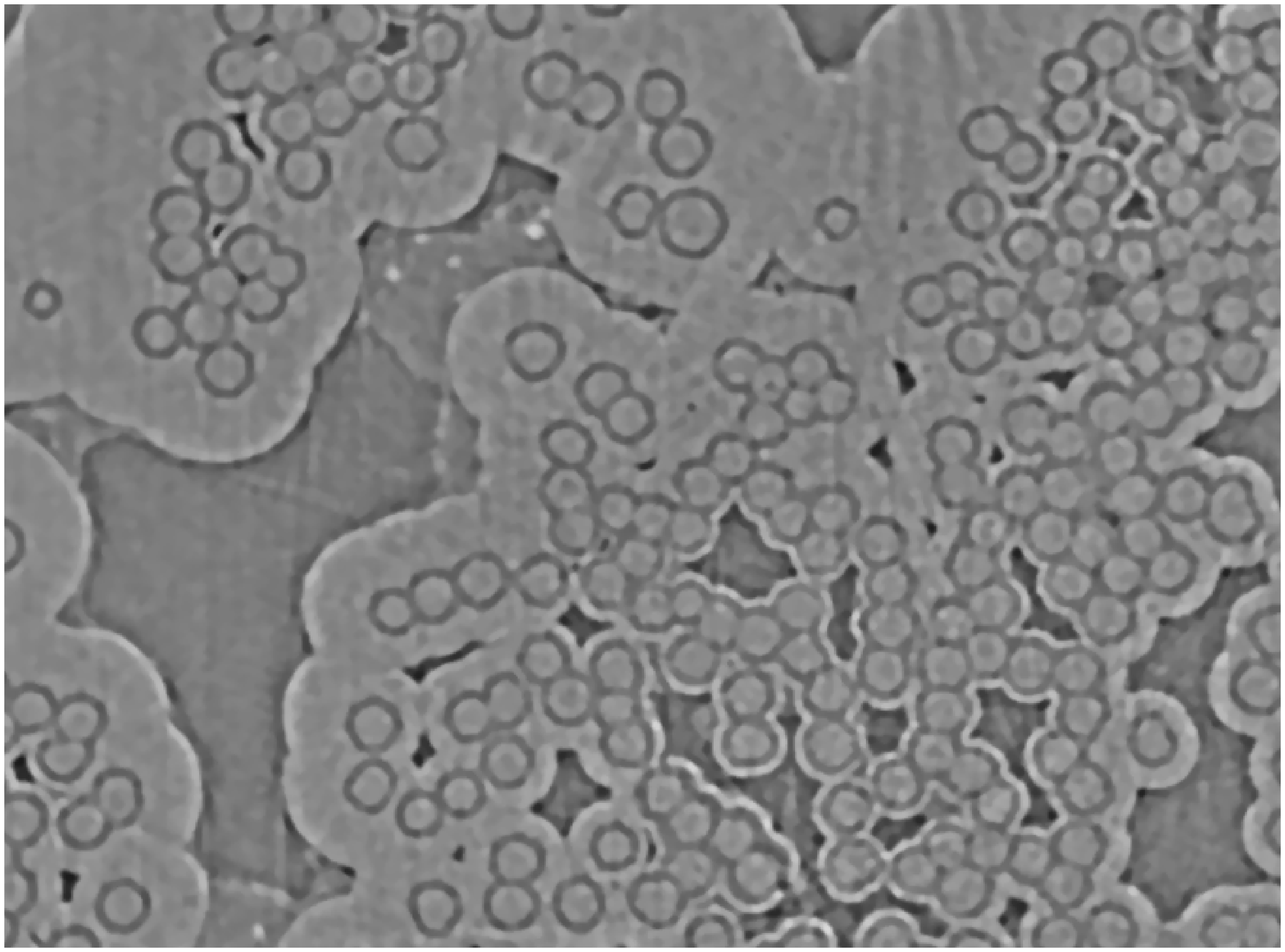, width=\imwidth}} \\
\end{tabular}
\end{center}
\vspace{-0.1in}
\caption{\label{fig:MBIRvsGridrec} A single (cropped) cross section from a 3D reconstructed volume of a carbon fiber data set acquired using a synchrotron X-ray instrument (adapted with permission from \cite{AdityaICASSP14}). 
Due to mis-calibrated detectors, direct use of the  analytic \textit{gridrec} algorithm results in a  reconstruction with ring artifacts. 
However, the use of a specially designed MBIR method suppresses the  ring artifacts while preserving detail and reducing noise. 
}
\end{figure}

\begin{figure}[!h]
\begin{center}
\epsfig{figure=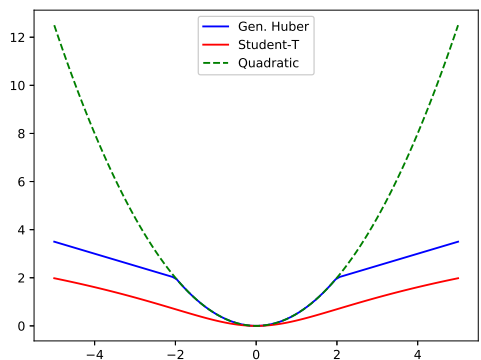, width=3in}
\caption{\label{fig:penalty} Penalty functions used for robust model-based image reconstruction. 
In several scientific CT applications the measurements can be corrupted  by outliers due to strong diffraction from crystalline samples, and gamma-ray/X-ray/neutron strikes on the detector in addition to detector noise.  
By using penalty functions based on heavy tailed probability density functions such as the generalized Huber function 
or the student-T 
function instead of the conventional quadratic function (dotted line) for the data-fidelity term, it is possible to obtain high-quality reconstructions with minimal pre/post-processing.}
\end{center}
\vspace{-0.3in}
\end{figure}

Finally, new forward models have also been formulated to address the challenge of  outliers due to gamma/X-ray/neutron strikes and spurious scatter due to Bragg diffraction when imaging samples that contain single-crystal domains. 
Because it is complicated to explicitly develop a physics-based model for such data, researchers have used new data-fidelity terms based on heavy tailed distributions for the $l$ in \eqref{eq:mbir_master} including the generalized Huber function \cite{VenkatBF15,AdityaICASSP14} and the student-T \cite{kazantsev2017novel} function (see Fig.~\ref{fig:penalty}) in the MBIR framework. 
Specifically, new forward models of the form  
\begin{eqnarray}
\label{eq:OutlierForwMod}
l(y;x) &=& \frac{1}{2} \displaystyle \Gamma((y-Ax)\sqrt{W})
\end{eqnarray}
where $\Gamma: R^{M} \rightarrow R$, $\Gamma(e)  = \displaystyle\sum_{i=1}^{M} \gamma(e_i)$, and $\gamma$ is of the form shown in Fig.~\ref{fig:penalty} have been used in \cite{VenkatBF15,AdityaICASSP14,kazantsev2017novel}.
While it is more complicated to find a minimum of the resulting cost-function, such an algorithm can further improve image quality compared to baseline MBIR algorithms based on a quadratic data-fidelity and analytic reconstruction algorithms.

\subsection{Time-Resolved CT \label{sec:mbir_4d}}
\vspace{-0.1in}
\begin{figure}[!htbp]
\vspace{-0.1in}
\begin{center}
\includegraphics[scale=0.52,trim=0cm 8cm 2cm 0cm,clip]{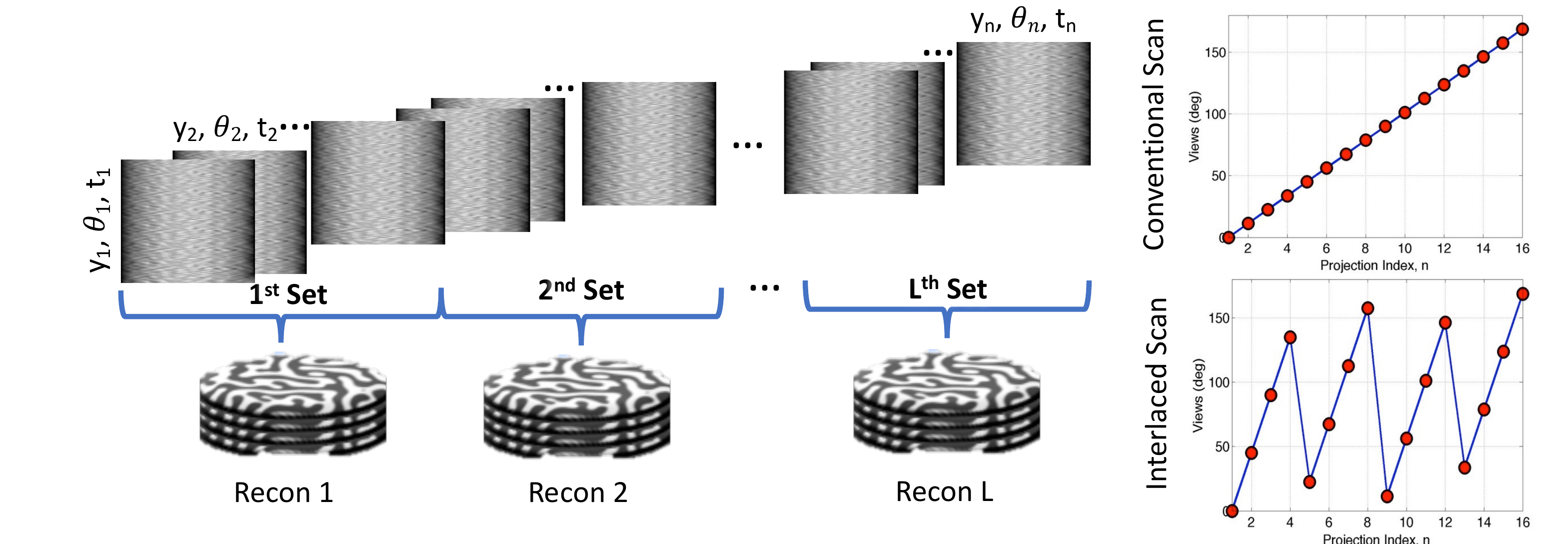}
\end{center}
\vspace{-0.3in}
\caption{\label{fig:4dschematic}
Illustration of the principle underlying 4D-CT of samples that are varying continuously based on the geometry of Fig.~\ref{fig:diff_inst} (a).
In this case, each measured projection image ($y_i$) corresponds to a specific orientation ($\theta_i$) and duration of time depending on the exposure and frame-rate of the detector.
The projections can be acquired by gradually orienting the sample in a 180 degree range or by using an interlaced scan as shown in the plots.  
In theory, each projection corresponds to a different state of the underlying 3D object. 
However, in practice, reconstruction algorithms are designed by assuming small variations, grouping collections of projection images and obtaining a 3D reconstruction corresponding to each collection.}
\vspace{-0.25in}
\end{figure}

SCT is also used to image the temporal dynamics of an object that is undergoing change in response to external stimuli such as varying temperature and pressure. 
This mode of imaging, commonly known as 4D-CT, is used to image the evolution of samples in 3D with respect to time.
In materials science, 4D-CT is used to study dynamic phenomena such as
\ifarxiv
solidification \cite{Gibbs2015Dendrites},
\else
solidification,
\fi 
\ifarxiv
phase transformations \cite{Aagesen2011Coarsening},
\else
phase transformations,
\fi
\ifarxiv
crack formation \cite{Bale20134DCracks}, 
\else
crack formation,
\fi
and
\ifarxiv
battery degradation \cite{Ziesche20204Dbattery,Pietsch2016PhaseCTBattery,Heenan2018NanoBattery}.
\else
battery degradation.
\fi
4D-CT is performed using a variety of radiation sources that include  
\ifarxiv
X-rays \cite{Ziesche20204Dbattery},
\else
X-rays,
\fi
\ifarxiv
electrons \cite{Kwon20104DElec},
\else
electrons,
\fi
and  
\ifarxiv
neutrons \cite{Totzke20174DNeutron},
\else
neutrons,
\fi
at resolutions ranging from nanometer to micron length scales mostly using the set-up of the type in Fig.~\ref{fig:diff_inst} (a). 
One way in which 4D-CT has been performed is by subjecting the material to the desired stimuli (like a certain pressure), acquiring a conventional CT for that specific stimulus point and repeating the process for different stimuli
\ifarxiv
stimuli \cite{Ziesche20204Dbattery,Pietsch2016PhaseCTBattery,Heenan2018NanoBattery}.
\else
stimuli.
\fi
The measurements corresponding to each CT scan is then reconstructed into a single 3D volume of the 4D reconstruction. 
Indeed, the 3D MBIR methods of section \ref{sec:mbir_3d} can be directly applied to these scenarios in-order to reduce the time-required to collect the data for a single reconstruction. 
However, if the goal of the study is to perform \textit{in-situ} imaging of the dynamics of rapidly changing material properties, then the overall problem becomes significantly more challenging. 

Fig.~\ref{fig:4dschematic} illustrates a data-acquisition scheme for 4D-CT of a sample which is changing continuously in the course of the measurement.
In order to conduct in-situ 4D-CT, the sample is rotated continuously about a single axis and the projection data is measured using a low-exposure setting on the detector in order to reduce motion blur. 
Each measurement corresponds to the projection of the sample at a certain time and orientation with respect to the incident beam.
Obtaining a 3D reconstruction corresponding to each time point is not possible because we only have a single projection image corresponding to that state of the sample. 
Therefore, algorithms designed for 4D-CT of continuously varying samples have used different strategies to acquire and process the data. 
The most-common approach is to group the data corresponding to a few orientations (see Fig.~\ref{fig:4dschematic}) and perform a 3D-CT reconstruction for each set with the implicit assumption that the sample does not change in the time window corresponding to each set. 
However, because of the wide-spread use of analytic reconstruction algorithms,  there has been a tendency to believe that each set need to contain measurements that cover a full angular range (typically $180^{\circ}$) and that a large number of such measurements (of the order of a few $1000$ for typical detectors) are required to obtain high-fidelity reconstructions. 
While some of these challenges can be overcome in specific situations (for example - the imaging of samples that are exhibiting a periodic motion \cite{walker2014vivo}),  
the overall use of analytic reconstruction approaches has limited the application of 4D-CT to the imaging of relatively slow processes. 
\newcommand{\denwidth}{1.7in}
\begin{figure}[!htbp]
\vspace{-0.025in}
\begin{center}
\begin{tabular}{ccc} 
\mbox{\epsfig{figure=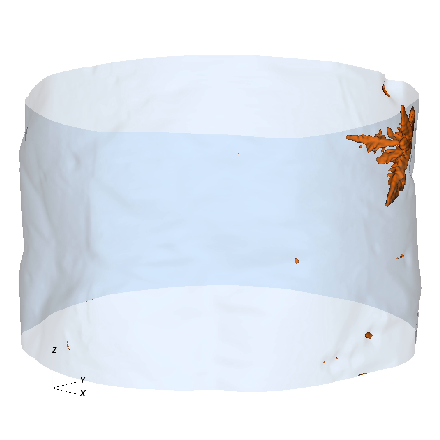, width=\denwidth}} &
\mbox{\epsfig{figure=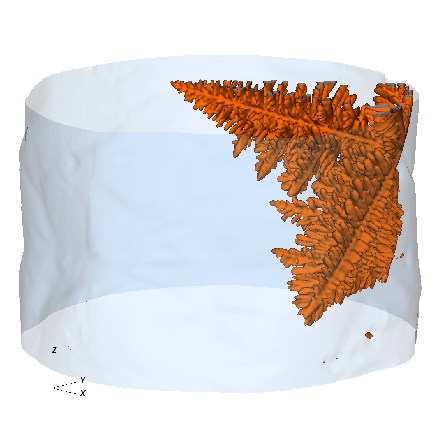, width=\denwidth}} \vspace{-0.3in} & 
\mbox{\epsfig{figure=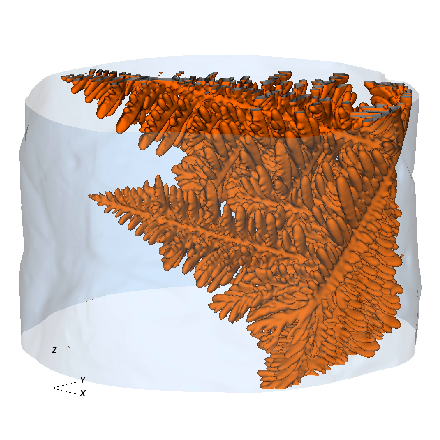, width=\denwidth}} \\
\end{tabular}
\end{center}
\vspace{-0.1in}
\caption{\label{fig:Dendrites4D}
4D-CT of growing dendrites in a slowly cooled Al-Cu alloy measured using synchrotron-based X-ray CT using the time-interlaced MBIR  \cite{Gibbs2015Dendrites}. 
A few reconstructed volumes from the 4D reconstruction that contain 3D frames every $1.8$ seconds at a voxel resolution of approximately $1 \mu$m, showing that it is possible to obtain high spatio-temporal resolution using MBIR approaches.
}
\vspace{-0.25in}
\end{figure}
In order to address the challenge of obtaining high-fidelity in-situ 4D-CT, researchers have adapted the MBIR methods discussed in section \ref{sec:mbir_3d} combined with synergistic changes in the way the data is acquired \cite{kaestner2011spatiotemporal}. 
One prominent approach is the time-interlaced MBIR (TIMBIR) \cite{AdityaTIMBIR} method in which the data is acquired such that the orientation angles ($\theta_{i}$ in Fig.~\ref{fig:4dschematic}) corresponding to each set are sparsely spaced over an angular range of $180$ degrees and interlaced with the angles in the other sets.
This interlacing offers increased measurement diversity compared to a standard accelerated/sparse-view system in which the collection of $\theta_{i}$ in each set is the same.  
Using a forward model that is similar to \eqref{eq:OutlierForwMod} for the data-fidelity term in \eqref{eq:mbir_master}, the authors proposed a regularizer of the form 
\begin{eqnarray}
r(x;\beta_s,\beta_t) = \beta_s \displaystyle \sum_{l=1}^L\sum_{\{i,j\} \in \chi_s} w_{ij}\rho_s(x_{l,i} - x_{l,j})
        + \beta_t \displaystyle \sum_{i}\sum_{\{m,n\} \in \chi_t} w_{mn}\rho_t(x_{m,i} - x_{n,i}),
\label{eq:4dct_prior}
\end{eqnarray}
where $\beta_s$ and $\beta_t$ are the regularization parameters for spatial and temporal regularization respectively, $x_{l}$ corresponds to the 3D reconstruction for set $l$, $\chi_s$ and $\chi_t$ represent the sets for the pairs of voxel neighbors across space and time respectively, $w_{ij}$ and $w_{mn}$, are weights associated with voxel pairs which are set to be inversely proportional to the distance/time between the neighbors. 
Using the TIMBIR algorithm, it has been demonstrated (see Fig.~\ref{fig:Dendrites4D}) that it is possible to obtain high-fidelity 4D reconstructions while accelerating the scan by a factor of $32$ compared to what would have been possible if the traditional protocol was used \cite{AdityaTIMBIR}.
The use of regularizers that
exploit local spatio-temporal correlations 
to improve reconstruction fidelity (similar to \eqref{eq:4dct_prior})
are widely studied in the research 
\ifarxiv
literature \cite{Wu20124DTV, Ritschl2012AdSpTemPrior}.
\else
literature.
\fi

While the regularizer \eqref{eq:4dct_prior} that exploits local spatio-temporal correlations are useful, researchers have also developed more sophisticated regularizers based on deformation fields or motion models 
\ifarxiv
\cite{Zang2018OptPriorNewSmpl,Zang2019WarpProj4D,Hinkle2012DiffMotMod} 
\else
\cite{Zang2018OptPriorNewSmpl}
\fi
demonstrating that it is possible to further improve the quality of the MBIR approaches for 4D-CT.
Other approaches that enforce spatio-temporal sparsity by
relying on the similarity between
non-local image patches have also been 
\ifarxiv
explored \cite{Tian2011TempNLMeans}.
\else
explored.
\fi
In summary, the key advantage of using MBIR approaches is that the measured data can be grouped in several different ways (each set potentially corresponding to sparse, low-SNR and limited-view sets) and jointly used for reconstruction thereby enabling high-fidelity images at unprecedented temporal resolutions.

\subsection{Accelerating MBIR for Large Data-Sets \label{sec:mbir_acc}}
While methods to accelerate MBIR depend on the specifics of the forward model, much of the recent research is focused on the conventional parallel-beam CT using models of the form \eqref{eq:3dforward} and \eqref{eq:mrf_prior} because of its wide-spread use across SCT applications. 
Broadly, the solutions to the MBIR cost-function based on \eqref{eq:3dforward} and \eqref{eq:mrf_prior} for conventional CT can be categorized into parallel update methods (like gradient decent) and sequential update methods (like coordinate-descent \cite{SaBo92}).
In either case, such algorithms are computationally expensive since it involves a large number of forward projection (multiplication by $A$) and back projection (multiplication $A^t$) operations which are typical in any iterative solution to the cost-function minimization.
For 4D-CT, this problem is compounded since  the number of views can be an order of magnitude larger than for 3D-CT. 
The computational complexity of the forward and backward projection operations also increases with the size of the reconstructed volumes. 
Thus, it becomes difficult to obtain real-time feedback
on the success of an experiment due to the long computational times of MBIR algorithms.
Furthermore, tuning of regularization and other free parameters becomes tedious in the absence of fast reconstructions. 
Thus, it is important to speed up MBIR algorithms for increasing their adoption for SCT.

One popular approach to speed up MBIR is to use novel optimization techniques
that speed up algorithmic convergence. 
The techniques to improve algorithmic convergence is typically dependent on the choice of the optimization algorithm used for reconstruction.
Multi-resolution approaches
\ifarxiv
\cite{Oh2006Multigrid,Oh2005Multigrid,Frese1999Multiscale,VenkatHAADF13,AdityaTIMBIR} 
\fi
use reconstruction at coarser resolution scales to initialize reconstruction at finer resolution. 
Such approaches are typically used to improve convergence of iterative coordinate descent (ICD) algorithms since ICD has poor low-frequency convergence \cite{VenkatHAADF13,AdityaTIMBIR}.
Another approach to speed up MBIR is  to use high-performance compute clusters (HPC) for distributed parallel computing \cite{wang2017massively,Bicer2015TLarScPar,Wang2016SVSIGPLAN}.  
This approach relies on modifications to existing optimization algorithms that enable distributed computation on super-computing high-performance compute clusters (HPC). 
In \cite{AdityaTIMBIR}, an approach to distributed parallel 4D-CT is presented where several 2D slices of each 3D volume over multiple time frames are reconstructed in parallel. 
However, this particular strategy of parallelizing over several 2D slices provides limited speed-up improvements since reconstruction of each slice is computationally expensive. 
Recently, algorithmic approaches for parallel reconstruction of voxels within each slice have also been proposed. 
These large scale parallelization approaches have led to a dramatic acceleration of reconstruction times of large volumes by distributing the computation across thousands of cores  - enabling reconstructions of size
$2160 \times 2560 \times 2560$ in about $24$s using $146880$ cores of a HPC cluster \cite{wang2017massively}. 

\ifarxiv
Finally, we note that there have been significant efforts to build use-friendly open-source software that enable application scientists to get access to advanced reconstruction algorithms. 
Software packages including TomoPy \cite{Tomopy14}, the ASTRA-toolbox \cite{vanAarleASTRA16,AstraGPU11,AstraUltramic15}, TIGRE \cite{biguri2016tigre}, ODL \cite{odl}, CIL \cite{jorgensen2021core}, svMBIR \cite{svmbir21}, pyMBIR \cite{pymbir}, and ToMoBAR \cite{tomobar20}, use powerful parallel-computing resources and are able to reconstruct large SCT data sets in time frames that permit experimentation, thereby accelerating the transition of advanced algorithms from research code to being used for experimental scans. 
The continued maintenance and development of such software packages will play a critical role in enabling the widespread adoption of advanced reconstruction algorithms. 
\fi

\subsection{
Regularization Parameters for MBIR
}

The eventual goal of SCT scans is either to discover new features of scientific relevance (like the appearance of a crack from a material under stress) or to perform a measurement (such as porosity of a manufactured part) from the reconstructed volumes. 
The choice of regularization function and their associated parameters has a significant impact on the quality (noise, resolution) of reconstructions obtained using the MBIR approach. 
Therefore, algorithms to automatically choose the regularization parameters that produce images that maximize the performance of the end-goal will be impactful.
However, due to the diversity of measurement scenarios, samples scanned and resolution values, the task of automatically choosing regularization parameters even for a fixed choice of regularization function of the form in \eqref{eq:mrf_prior} is challenging. 
Furthermore, the regularization parameters are often dimensionless quantities which do not have a straightforward interpretation for end-users of SCT instruments, making them complicated to set in an intuitive manner.  

While some general approaches for setting the regularization parameters in the context of model-based reconstruction have been proposed (\cite{ramani2008monte} and references therein), they have not been widely adapted for SCT.
These approaches can be broadly categorized as methods that require evaluation of multiple reconstructions (L-curve, generalized cross validation etc.); those that set the value based on ``balancing'' the data-fidelity and regularization terms; and Bayesian methods that jointly estimate the regularization parameter and reconstructions 
\ifarxiv\cite{saquib1998ml}\fi.
The above approaches are not guaranteed to produce reconstructions of the best quality matched to the task for which SCT is being carried out.
The extensions of these methods for applications like 4D-CT and hyper-spectral CT has also not been explored. 
Current adaptations of MBIR for SCT have mainly relied on an empirical choice of parameters in order to attain some desirable visual image quality (a certain level of noise, sharpness of edges etc.). 
Typically the parameters are varied, a few slices from the 3D volume are reconstructed in order to manage the computational complexity, and for each choice of parameters the reconstruction is evaluated sometimes based on a predefined metric \cite{allner2019metric}.
In summary, the regularization parameters for SCT applications have largely been set in an empirical manner, making the automated choice of these parameters an important future research direction. 

\section{Deep Learning-Based CT Reconstruction \label{sec:dl}}






\begin{figure}
\begin{center}
\includegraphics[scale=0.5,trim=0cm 0.5cm 6cm 0cm,clip]{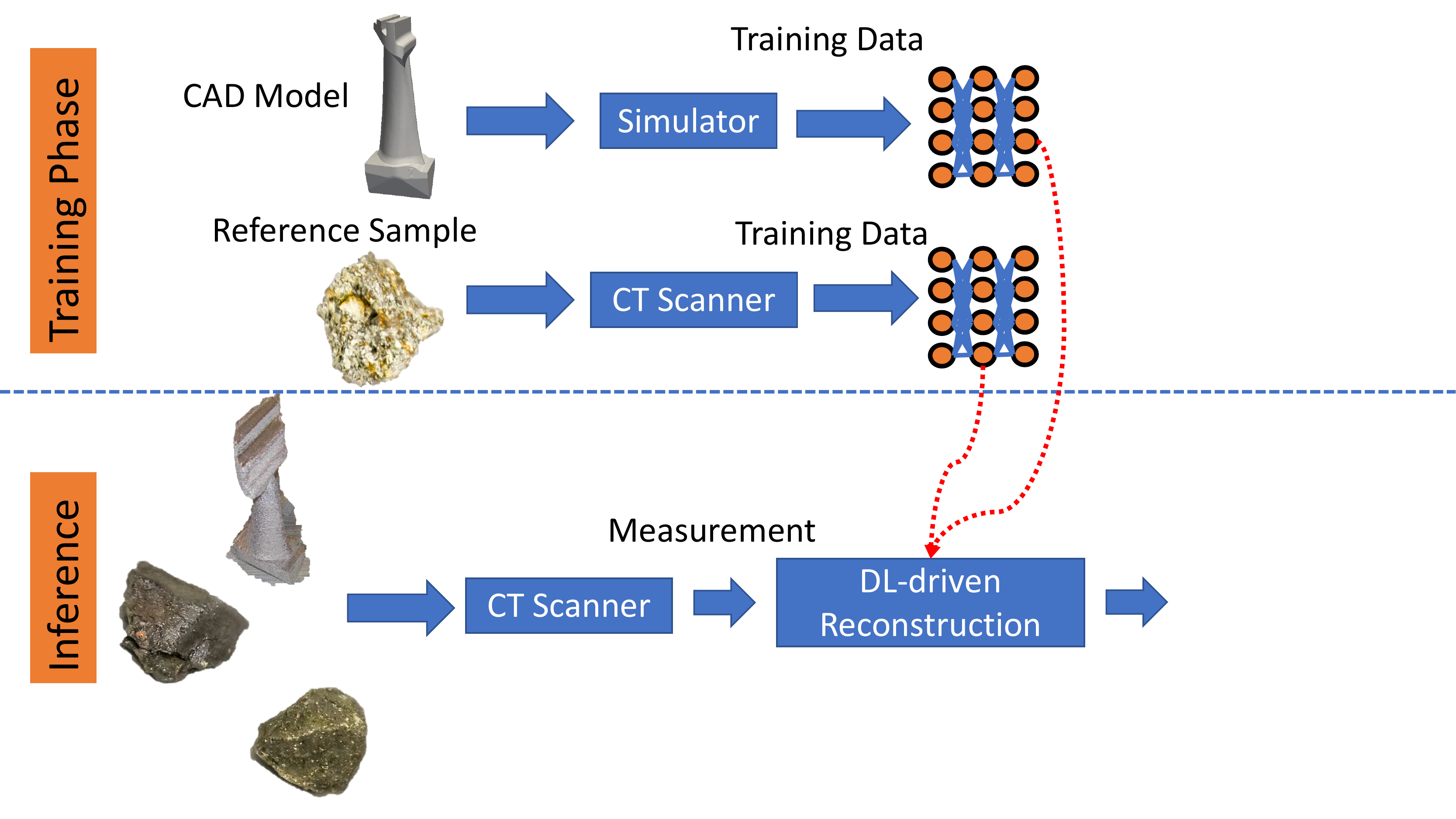} 
\end{center}
\vspace{-0.3in}
\caption{\label{fig:overallpicture} Illustration of approaches used for deep-learning based SCT.
The training data is either generated from a reference scan of a representative sample or from a computer-aided design based  model when appropriate.
Due to the diversity of samples to be measured using SCT, it is challenging to train a single general purpose neural network that works across a wide variety of samples; hence current research has focused on training neural networks for a specific collection of samples. 
}
\end{figure}
Deep learning (DL) based algorithms have recently been developed in order to address several challenges that occur in the context of CT reconstruction \cite{wang2020deep}.   
These algorithms can be broadly categorized into iterative and non-iterative approaches \cite{mccann2017convolutional}. 
Iterative approaches to DL are based on explicitly deriving the iterative updates that result from solutions to the MBIR formulations of Section.\ref{sec:mbir_3d} and then replacing certain blocks with trainable deep neural networks (DNN) in order to obtain high-quality reconstructions 
\ifarxiv
(including methods categorized as loop-unrolling algorithms \cite{monga2021algorithm} 
or plug-and-play priors \cite{VenkatPlugPlay13,suhaspnp16})
\fi
.
In contrast, non-iterative approaches are based on training a DNN to learn to pre-process the measurements followed by the use of a conventional analytic reconstruction algorithm. 
Alternately, non-iterative DL (NIDL)-based algorithms can be applied in the reconstruction domain by designing a DNN to learn to suppress common artifacts that occur when using analytic reconstruction algorithms\ifarxiv
~\cite{kang2017deep,chen2017low,kang2018deep,jin2017deep,han2018framing,liu2020tomogan,ziabari20182,ziabari2020IMECE,venkatMLST20}
\fi.
In the context of SCT, NIDL-based approaches have garnered significant interest because of their low computational complexity at inference time and simple portability to graphics processing units that can enable fast reconstructions for the large data-sets from SCT instruments.   

While NIDL-based reconstruction has been widely explored in the context of applications such as medical X-ray CT, there are several challenges in adapting these methods for SCT applications. 
First, SCT instruments are used to scan a wide variety of samples and therefore the training data for DNNs has to be chosen carefully.
From an algorithm designer's perspective, it may be impossible to obtain sufficient data in order to train a single generic DNN that can be applied for any sample that has to be imaged under different measurement scenarios (geometry, number of views, SNR etc.). 
Furthermore, obtaining high-quality reference data-sets to train neural networks can be time-consuming/expensive, and therefore in SCT applications we might only be able to make a very small number of reference measurements. 
Finally, the central problems associated with SCT are high-dimensional involving 3D spatial and 4D spatio-temporal reconstructions, leaving open the question of how to effectively adapt popular DNN architectures that are designed for 2D imaging applications such as the U-Net\cite{ronneberger2015u} and DnCNN \cite{zhang2017beyond} to the multi-dimensional case. 

\begin{figure}
\vspace{-0.1in}
\begin{center}
\includegraphics[scale=0.9,trim=6.5cm 6.6cm 0cm 0cm,clip]{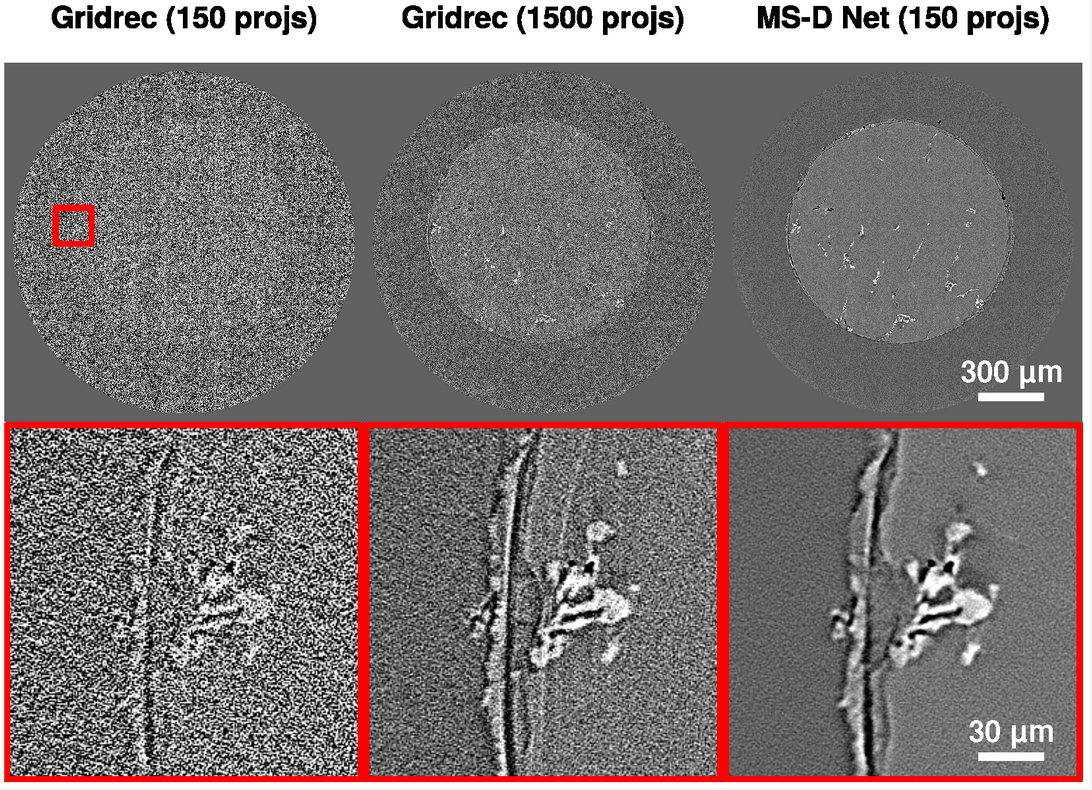}
\end{center}
\vspace{-0.25in}
\caption{\label{fig:dl_recon_3d} Illustration of the use of DL-based reconstruction for sparse-view and low-SNR synchrotron X-ray CT data-set of an aluminum sample (reproduced from \cite{pelt2018improving}). 
The figure shows a single cross-section from a 3D reconstruction using an analytic reconstruction algorithm (Gridrec) and a DL-based reconstruction based on the mixed scale dense neural network. 
Notice that using one-tenth the number of measurements typically made, the deep neural network based approach is able to produce higher-quality reconstruction compared the analytic reconstruction approach.} 
\end{figure}
Supervised NIDL algorithms have been developed for SCT mainly in the context of scanning collections of similar samples, 4D-CT, or those situations for which a 3D model is available as shown in Fig.~\ref{fig:overallpicture}.
If the goal of an experiment is to accelerate the measurement of a large collection of similar samples (a set of rocks, additively manufactured parts etc.), then one or more of the samples can be scanned in a manner such that we can obtain training data for a DNN that is designed to obtain high-quality reconstruction from sparse/limited-view/low-SNR data. 
A variety of different approaches to obtain training data, choice of network architecture and loss-functions have been explored in the context of developing NIDL for SCT. 
For example, in \cite{yang2018low} a NIDL approach was developed in order to obtain high-quality reconstruction from low-dose X-ray CT data. 
A few pairs of low-dose and high-dose projection images were measured,  followed by the use of a encoder-decoder based DNN trained using a Wasserstein and perceptual loss function in order to map between the low-dose and high-dose projection data. 
Since the data is extremely limited, the acquired images were split into smaller patches and  data-augmentation techniques were used in order to stably train the DNN which has a very large number of parameters. 
The trained network was then used to obtain high-quality CT reconstruction by processing new low-dose measurements using the DNN followed by the use of a analytic reconstruction approach - both steps which can be performed rapidly for the large data sets encountered in this application. 
Another NIDL approach for SCT has involved measuring a complete low-dose and regular-dose CT scan from a reference sample and reconstructing them using an analytic reconstruction algorithm.
This data  
is used to train a DNN to map between the low-dose and high-dose 3D reconstruction. 
One challenge in such approaches is that there may only be a single 3D volume pair in order to train the network.
In order to address this limited-data challenge, methods of splitting the 3D volume into smaller patches combined with data-augmentation methods such as flipping and rotation are used to increase the size of the training set \cite{liu2020tomogan,ziabari2020IMECE}. 
In contrast to this approach, a new neural network architecture - mixed-scale densenet (MS-DNet)  -  was proposed in \cite{pelt2018improving, pelt2018mixed}
which has an order of magnitude fewer parameter than other popular DNN architectures \cite{ronneberger2015u,zhang2017beyond} and a large-receptive field making it a strong candidate for the limited training data encountered in SCT applications. 
Furthermore, in order to exploit the 3D structure of the data, instead of training a fully 3D DNN, the works in \cite{pelt2018improving,liu2020tomogan, pelt2018mixed} have used a 2.5D strategy \cite{ziabari20182} 
where the input to the neural network is a collection of adjacent slices (modeled as channels) and the target output is a single image.  
The overall approach of using NIDL based on making one high-quality volumetric reference, has demonstrated that it is possible to accelerate acquisition and yet obtain high-quality 3D reconstructions from extremely sparse-view data (see Fig.~\ref{fig:dl_recon_3d}) which can be used to dramatically improve the throughput of SCT instruments at shared facilities \cite{venkatMLST20}.
However, the generalization ability of different approaches which drives the need for measuring new training data sets remains an open question. 

NIDL-based algorithms are also being used when a computer-aided model (CAD) of a part to be scanned is available (typical in the case of additively manufactured objects) \cite{ziabari2020IMECE}.
The goal of these studies is to reduce measurement time for high-resolution CT and to obtain high-quality reconstructions from samples that can potentially introduce non-linearities such as beam-hardening into the measurement. 
For example, in \cite{ziabari2020IMECE} 
a framework was developed to generate training data for DNNs by simulating X-ray CT scans of a specific part in a polyenergetic X-ray micro-CT system.
In order to obtain realistic data, a simulator was developed that embeds various defects in the CAD model, accounts for complex phenomenon such as beam-hardening of the incident X-rays and generates the sparse, low-SNR data that occurs during an experimental CT scan.
Since the ground truth is known,  a 2.5D DNN was trained  
based on a conventional mean-squared loss function in order to suppress noise, streaks and beam-hardening artifacts that are encountered when using analytic reconstruction approaches.  
Once this network was trained, it was applied to experimental data of samples corresponding to the CAD model, demonstrating promising preliminary results in obtaining higher quality reconstructions from accelerated scans while enhancing the detectability of defects. 

Finally, self-supervised DL approaches have also been used for enhancing the quality of SCT reconstructions, especially when obtaining pairs of matched training data for supervised learning is not possible.
A recent example is~\cite{hendriksen2020noise2inverse} where a DNN  was trained from a single low-SNR experimental CT scan by splitting the acquired data in an effective manner.
Specifically by obtaining a pair of 3D reconstructions from sub-sets of the measured data, a DNN based on the MS-DNet \cite{pelt2018improving} was trained to map between pairs of these reconstructed slices. 
One trained, the same network can be applied to the entire noisy measurement in order to produce a high-fidelity reconstruction. 
The method, deemed Noise2Inverse, further expanded in in~\cite{hendriksen2021deep} and was successfully applied to static and dynamic micro-tomography as well as X-ray
diffraction tomography. Substantial reduction in acquisition time was achieved while reducing noise and maintaining image quality.
\color{black}
This type of self-supervised learning approach can be particularly useful for SCT applications such as neutron CT or lab-based X-ray CT where obtaining high-resolution scans are extremely time-consuming. 

\section{Conclusion \label{sec:concl}}

In this article we presented an overview of how advanced image reconstruction algorithms are enabling improvements in the performance of scientific CT instruments. 
These algorithmic advances are a powerful complement to the decades of advances in hardware technologies in being able to obtain high-fidelity images while enabling dramatic acceleration of the time required to make measurements. 
While the research highlighted in this article has demonstrated the potential for algorithm-driven approaches for SCT instruments, there are still several open questions including the choice of regularization function and parameters for MBIR methods, algorithmic and computation driven acceleration for larger data-sets, and the further investigation of the choice of architecture, parameters and loss-functions for deep learning-based reconstruction. 



\bibliographystyle{IEEEtran}
\bibliography{tomo}


\end{document}